\documentclass[sigconf, authorversion]{acmart}
\AtBeginDocument{%
  \providecommand\BibTeX{{%
    \normalfont B\kern-0.5em{\scshape i\kern-0.25em b}\kern-0.8em\TeX}}}

\copyrightyear{2023} 
\acmYear{2023} 
\setcopyright{acmlicensed}\acmConference[SIGSPATIAL '23]{The 31st ACM International Conference on Advances in Geographic Information Systems}{November 13--16, 2023}{Hamburg, Germany}
\acmBooktitle{The 31st ACM International Conference on Advances in Geographic Information Systems (SIGSPATIAL '23), November 13--16, 2023, Hamburg, Germany}
\acmPrice{15.00}
\acmDOI{10.1145/3589132.3625601}
\acmISBN{979-8-4007-0168-9/23/11}

\usepackage{import,transparent}
\usepackage{nicefrac}
\usepackage{makecell}
\usepackage{algorithm}
\usepackage{algorithmic}
\usepackage{subcaption}
\usepackage{xspace}
\usepackage{pgfplots}
\usepackage{pgfplotstable}
\usepackage{booktabs}
\usepackage{multirow}
\usepackage{wrapfig}
\usepackage{paralist}
\usepackage{pifont}
\usepackage{balance}
\usepackage[aboveskip=1ex, belowskip=-1ex]{caption}
\usepackage{tikz}
\usetikzlibrary{shapes,arrows, positioning, shapes.geometric,fit,backgrounds,calc}
\usetikzlibrary{patterns}
\tikzset{every mark/.append style={mark size=1pt}}
\usetikzlibrary{pgfplots.colorbrewer}
\usepgfplotslibrary{groupplots}

\pgfplotsset{height=3.5cm, 
    label style={font=\footnotesize},
    tick label style={font=\footnotesize}, 
    title style={font=\footnotesize, yshift=-1.2ex},
    ylabel shift=-1ex,
    xlabel shift=-0.8ex,
    axis lines*=left,
    grid=major,
    grid style={line width=.1pt, draw=gray!20},
    legend style={
        font = \scriptsize,
        fill = white!90, 
        draw opacity = 1,
        text opacity = 1,
        very thin,
        legend columns = 3,
        draw=none, 
    },
    compat=newest}

\pgfplotsset{
    barPlotStyle/.style={
        enlarge y limits = 0,
        enlarge x limits =.1,
        x tick label style={rotate=45, anchor=east},
	    cycle list/.style={
            colormap/Paired,
            cycle list/Paired,
        },
        every axis plot/.append style={fill},
        label style={font=\footnotesize},
    }
}

\pgfplotsset{
    linePlotStyle/.style={
    every axis plot/.append style={line width=1pt},
    legend style = {legend columns = -1},
    colormap/Paired,
    cycle list/Paired,
    cycle multiindex* list={
        mark list*\nextlist
        Paired\nextlist
        linestyles\nextlist
    },
    label style={font=\footnotesize},
}}

\usepackage{scalerel}
\def\msquare{\mathord{\scalerel*{\Box}{gX}}}

\usepackage{fontawesome5}
\usepackage{tikz}
\usetikzlibrary{shapes, shapes.geometric, shapes.arrows, arrows, patterns, positioning, calc, arrows.meta, bending}
\usetikzlibrary{mindmap,trees}
\definecolor{mygreen}{rgb}{0.553,0.682,0.063}
\definecolor{myblue}{rgb}{0.0,0.208,0.376}
\definecolor{mygray}{rgb}{0.906,0.906,0.906}

\definecolor{darkblue}{rgb}{0.0,0.0,0.8}
\definecolor{mydarkgreen}{rgb}{0.0,0.5,0.0}
\definecolor{tablehighlight}{rgb}{0.0,0.5,0.0}

\newcommand{\ie}{i.\nolinebreak[4]\hspace{0.01em}\nolinebreak[4]e.\@\xspace}
\newcommand{\eg}{e.\nolinebreak[4]\hspace{0.01em}\nolinebreak[4]g.\@\xspace}

\newcommand{\sysname}{\emph{RapidEarth}\@\xspace}

\begin{document}



\title{RapidEarth: A Search-by-Classification Engine for Large-Scale Geospatial Imagery (Demo Paper)}

\author{Christian Lülf}
\affiliation{%
  \institution{University of Münster}
  \city{Münster}
  \country{Germany}
  \postcode{48153}
}
\email{christian.luelf@uni-muenster.de}

\author{Denis Mayr Lima Martins}
\affiliation{%
  \institution{University of Münster}
  \city{Münster}
  \country{Germany}
  \postcode{48153}}
\email{denis.martins@uni-muenster.de}

\author{Marcos Antonio Vaz Salles}
\affiliation{%
  \institution{Independent Researcher}
  \country{Portugal}
}
\email{msalles@acm.org}

\author{Yongluan Zhou}
\affiliation{%
  \institution{University of Copenhagen}
  \city{Copenhagen}
  \country{Denmark}
}
\email{zhou@di.ku.dk}

\author{Fabian Gieseke}
\affiliation{%
  \institution{University of Münster}
  \city{Münster}
  \country{Germany}
  \postcode{48153}}
\email{fabian.gieseke@uni-muenster.de}


\begin{abstract}
Data exploration and analysis in various domains often necessitate the search for specific objects in massive databases. A common search strategy, often known as \emph{search-by-classification}, resorts to training machine learning models on small sets of positive and negative samples and to performing inference on the entire database to discover additional objects of interest. While such an approach often yields very good results in terms of classification performance, the entire database usually needs to be scanned, a process that can easily take several hours even for medium-sized data catalogs.
In this work, we present \sysname, a geospatial search-by-classification engine that allows analysts to rapidly search for interesting objects in very large data collections of satellite imagery in a matter of seconds, without the need to scan the entire data catalog. \sysname embodies a co-design of multidimensional indexing structures and decision branches, a recently proposed variant of classical decision trees. These decision branches allow \sysname to transform the inference phase into a set of range queries, which can be efficiently processed by leveraging the aforementioned multidimensional indexing structures. 
The main contribution of this work is a geospatial search engine that implements these technical findings. 

\end{abstract}

\ccsdesc[500]{Information systems~Search engine indexing}
\ccsdesc[300]{Computing methodologies~Classification and regression trees}

\keywords{search engine, decision trees, classification, index structures}


\maketitle


\section{Introduction}
\label{sec:introduction}
The tasks of geospatial imagery analysis and exploration are confronted with an ever-increasing challenge of managing and mining vast amounts of data. This surge in data volume is notably attributed to the advancement in satellite technology and the frequency of image capture~\cite{songnian2016biggeodata}. Given this growth in data, there is a pressing need for tools that empower users to navigate through these massive amounts of data quickly.

In response to this need, a very recent and promising approach has emerged that comprises a variant of decision trees called \textit{decision branches}, bundling together machine learning inference with database index structures~\cite{lulf2023fast}. This method enables the fast identification of objects of interest within large data catalogs in a matter of mere seconds. 
However, it remains unclear how such an approach can be integrated into a seamless geospatial search experience. Ideally, departing from only a few labeled images as a reference point, users should be able to rapidly locate similar objects in satellite imagery---for instance, identifying areas of forest degradation within the Amazon.

In this paper, we present \sysname, a system that leverages decision branches to construct a fast and effective geospatial search engine for satellite imagery. \sysname is the first-ever prototype to integrate decision branches in an end-to-end application, demonstrating their applicability in a large-scale setting. \sysname features a user-friendly graphical web interface where users can effortlessly navigate across aerial imagery 
and interactively define their search queries by pointing and clicking on a map.

So far, previous geospatial search engines on image similarity have traditionally depended on \emph{k}-nearest neighbor approaches, which return the \emph{k} closest images relative to the chosen input image \cite{zhang2022,KEISLER2019visualsearch,icde2020,icmr2012}. Despite their fast query time (\ie, time to produce results once a user query is issued), query results may lack comprehensiveness due to inherent limitations of the underlying methods: The returned images are restricted to the top \emph{k} neighbors, and the query intent can be defined by only a single item. By contrast, using decision branches, we can transform a search task into a binary classification problem, which allows us to train a machine learning model using multiple objects of interest (\eg, class label 1) and even include dissimilar objects that should be excluded from the query results (\ie, class label 0), enabling more precise delineation between the objects of interest and the rest. Thereby, we expect query results to be more precise and complete in comparison to previous search engines. By coupling the search model with index structures, we ensure that the query response time remains competitive to \emph{k}-nearest-neighbor-based search solutions.

\noindent\textbf{Contributions.} Key contributions of this work are:
\begin{enumerate}[(i)]
    \item We present a prototype of our \sysname search engine that implements a search-by-classification approach with decision branches for an aerial data set (see Figure~\ref{fig:screenshot}). The prototype is publicly available at \url{https://web.rapid.earth}.
    \item Through a set of demonstration scenarios, we illustrate how \sysname can be used to locate target objects, such as solar panels, in rich satellite imagery. The selected demonstration scenarios will be the basis for discussions with the attendees regarding the underlying decision branches technique and its implementation in \sysname.
    \item We also provide the source code that allows other researchers to (re)use the search engine for their own data and use cases. The code repository is publicly available at \url{https://github.com/decisionbranches/rapidearth}.
\end{enumerate}
The rest of the paper is structured as follows. Section~\ref{sec:approach} describes the workflow of our search engine. Section~\ref{sec:dataset} elaborates on the data set used for our prototype. We delve into the system architecture of \sysname in Section~\ref{sec:prototype}. Subsequently, Section~\ref{sec:demo} outlines the application of \sysname through a practical demonstration including the search of \emph{solar panels} over the country of Denmark. The paper ends with a short summary in Section~\ref{sec:summary}.
\section{Approach}
\label{sec:approach}
\begin{figure}[t]
    \centering

\tikzstyle{database}=[
    cylinder,
    cylinder uses custom fill, 
    cylinder body fill = green!20, 
    cylinder end fill = green!30,
    aspect = 1.5,
    shape border rotate = 90,
    minimum height = 3.0em,
    minimum width  = 2.5em,
    draw = gray,
    anchor=west,
    thick,
]
\tikzstyle{block}=[rectangle, 
	rounded corners,
    draw=black, 
	text=black,
	minimum width=1.0cm, 
	minimum height=1.0cm,
	color=black,
]

\tikzstyle{asset}=[rectangle,
    minimum width=0.1cm,
    minimum height=0.1cm,
    font = \Large,
    color=black,
]

\tikzstyle{arrowicon} = [pos=0.5,
    inner sep=0.05cm,
    color=black,
    fill=white,
]

\tikzstyle{index}=[
    isosceles triangle,
    isosceles triangle apex angle = 60,
    rotate = 90,
    draw = orange, 
	fill = orange!20,
    minimum size = 1.5em,
    anchor = apex,
]

\tikzstyle{arrow} = [->,-To, semithick, color=black]

\tikzstyle{customlabel} = [color=black]

\begin{tikzpicture}[
    font=\normalsize,
    scale=0.7,
    every node/.style={transform shape}
]

\node[database] (db) {};
\node[block, right=8mm of db, align=center, text width=20mm] (featext) {Feature Extractor};

\matrix[right=5mm of featext, column sep=-1mm, row sep=-3mm, inner sep=0.2mm] (vec1) {
    \node[asset] {\faIcon{square}}; &
    \node[asset] {\faIcon[regular]{square}}; &
    \node[asset] {\faIcon{square}}; & 
    \node[asset] {\faIcon[regular]{square}}; \\
};

\node[block, right=4mm of vec1, align=center, text width=20mm] (ibuilder) {Index Builder};

\node[index, below left=5mm and -6mm of ibuilder] (idx1) {};
\node[index, below left=5mm and -14mm of ibuilder] (idx2) {};
\node[index, below left=7mm and -9.8mm of ibuilder] (idx3) {};


\matrix[below=18mm of db, inner sep=0.2mm] (queries)
{
    \node[asset] {\faIcon[regular]{plus-square}}; & 
    \node[asset] {\faIcon[regular]{plus-square}}; &
    \node[asset] {\faIcon[regular]{minus-square}}; \\
    \node[asset] {\faIcon[regular]{minus-square}}; &
    \node[asset] {\faIcon[regular]{minus-square}}; &
    \node[asset] {\faIcon[regular]{minus-square}}; &\\
};

\matrix[below=19.2mm of featext, column sep=-1mm, row sep=-3mm, inner sep=0.2mm] (vec2) {
    \node[asset] {\faIcon[regular]{square}}; &
    \node[asset] {\faIcon{square}}; &
    \node[asset] {\faIcon[regular]{square}}; & 
    \node[asset] {\faIcon{square}}; \\
};

\node[block, right=15mm of vec2, align=center, text width=20mm] (cls) {Index-Aware\\Classifier};

\node[block, right=5mm of cls, align=center, text width=20mm] (rq) {Inference via\\ Range Queries};

\node[customlabel, right=3mm of rq, align=center] (obj) {Target\\Objects};

\node[customlabel, below=1mm of db] {Data};
\node[customlabel, below=0mm of vec1] {Features};
\node[customlabel, below right=6mm and -12mm of idx2] {Indexes};

\node[customlabel, below=0mm of queries, align=left] {User Query};

\node[customlabel, below=0mm of vec2] {Features};




\draw[arrow] (db.east) -- (featext.west);
\draw[arrow] (featext.east) -- (vec1.west);
\draw[arrow] (featext.south) -- (vec2.north);
\draw[arrow] (vec1.east) -- (ibuilder.west);
\draw[arrow] (ibuilder.south) -- (idx3.east);
\draw[arrow] (idx3.west) -- (cls.north);
\draw[arrow] (idx3.west) -- (rq.north);
\draw[arrow] (queries.east) -- (vec2.west);
\draw[arrow] (vec2.east) -- (cls.west);
\draw[arrow] (cls.east) -- (rq.west);
\draw[arrow] (rq.east) -- (obj.west);

\node[customlabel, anchor=west] at (-0.8, -1.3) {Offline Preprocessing};
\draw[gray, dashed] (-0.7, -1.6) -- (11.5, -1.6);
\node[customlabel, anchor=west] at (-0.8, -1.9) {Query Processing};

\end{tikzpicture}
    \caption{Overview of the search-by-classification workflow from~\cite{lulf2023fast} employed in our prototype.}
    \label{fig:pipeline}
\end{figure}
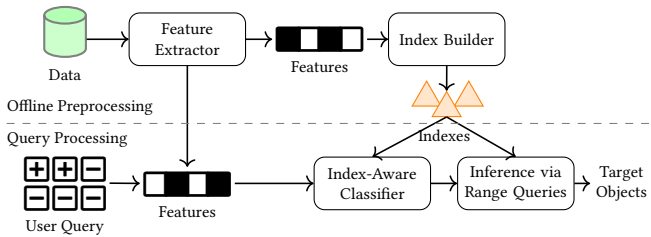

The workflow of processing a query with our search engine is sketched in Figure \ref{fig:pipeline}. The workflow is separated into two phases (1) \emph{Offline Preprocessing} and (2) \emph{Query Processing}. In the offline phase, executed prior to the search engine's launch, we pre-built the index structures needed to speed up the user query. Here, we first extract meaningful features from the data (in this case satellite imagery data) that capture some characteristics of the underlying data structures (\eg, colors, detected structures like streets, houses, etc.). How we extracted the features is described in Section \ref{sec:dataset} in more detail. Given the extracted features of all data, we start building the index structures which are based on k-d trees \cite{Bentley1975}.\footnote{ While various alternative index structures are available for efficiently handling range queries, we have chosen to implement k-d trees in our prototype due to its straightforward implementation.}

\begin{figure}[t]
    \centering
    \newcommand{\xfeature}{$x_1$}
\newcommand{\yfeature}{$x_2$}
\newcommand{\nodewidth}{6ex}
\newcommand{\nodeheight}{2ex}
\newcommand{\nodedistance}{4.5ex}
\newcommand{\leveldistance}{10ex}

\definecolor{boxcolor}{HTML}{20b49c}

\tikzstyle{treearrow} = [->,-To, shorten >=1pt]

\tikzstyle{treenode} = [
  draw,     
  shape=rectangle,
  color=black,
  solid, 
  minimum width=\nodewidth,
  minimum height=\nodeheight,
]

\tikzstyle{rareclass} = [
  draw,
  shape=rounded rectangle,
  color=black,
  fill=boxcolor!20,
  solid, 
  minimum width=\nodewidth,
  minimum height=\nodeheight,
]

\tikzstyle{nonrareclass} = [
  draw,     
  shape=rounded rectangle,
  color=black,
  solid, 
  minimum width=\nodewidth,
  minimum height=\nodeheight,
]
\begin{tikzpicture}[
    every mark/.append style={mark size=0.55ex},
    level distance=\nodedistance,
	level 1/.style={sibling distance=\leveldistance},
	level 2/.style={sibling distance=\leveldistance},
    node distance=2.2cm and 4cm,
    font=\scriptsize,
    scale=0.95,
    every node/.style={transform shape},
    ]
    \begin{axis}[
        at = {(0\linewidth,0)},
        width=0.5\linewidth,
        xmin=0, xmax=6,
        ymin=0, ymax=6,
        tick align=outside,
        xtick distance=1,
        ytick distance=1,
        tick label style={font=\scriptsize}, 
        xlabel={\xfeature},
        ylabel={\yfeature},
        label style={font=\footnotesize},
        legend style={
          at={(0.5,1.23)},
          anchor=north,
          legend columns=-1,
        },
    ]
     \addplot [
        scatter, 
        only marks,
        scatter src=explicit,
        scatter/classes={
            1={mark=*, fill=red, draw=red},%
            0={mark=square*, fill=black!70, draw=black, opacity=0.8}
        },
    ]
    table [
        x=x,
        y=y,
        meta=label,
    ] {%
    x  y label
    4.13353701507816 2.03743899035789 1
    2.66531709639346 3.02973229670563 1
    2.14397063539591 3.11875609246825 1
    1.99769850031271 2.39407401567857 1
    3.0945314721689 3.7702152126602 0
    3.39279501670053 4.2302340125135 0
    3.72092961493132 4.05642015776681 0
    2.95344837344581 4.83288178496143 0
    3.16344277501011 4.53800314164617 0
    4.36380975976508 3.15155061608832 0
    2.87857879055556 5.09160537837457 0
    1.7756870183853 3.76615461505307 0
    3.78486492736914 3.91420023248269 0
    1.8254799539577 4.30216675997238 0
    2.50918318446944 3.44569774324521 0
    1.28571823212024 4.46885064330025 0
    3.33161630291041 4.02787705347493 0
    2.85656885534383 4.65104803878442 0
    5.02464880501632 2.19006930756427 0
    2.42809461325994 4.10491315918182 0
    2.74721600669124 5.5032583962494 0
    2.73482726932059 2.12779667928195 0
    3.18598464597997 4.47619871388183 0
    3.48145256596113 4.04260164580873 0
    2.45211608392477 2.38148184276365 0
    2.17772221093714 4.07568830609915 0
    2.43815735523012 3.29095073208471 0
    3.01831874718029 3.50184210780366 0
    2.74598842862603 3.56095613306554 0
    2.97430966925448 4.78629981921542 0
    1.95475546031246 5.32104272294127 0
    3.18093448053639 4.62236345692378 0
    4.30535651558841 3.19410511986788 0
    3.91556974420768 3.72005587139902 0
    3.90483015224456 3.7645002618933 0
    2.06807899097868 2.13496399821754 0
    3.07915150009021 4.90510686299592 0
    3.88294775568109 3.77391846594387 0
    1.50763318228028 3.19213378633646 0
    2.56607920266634 1.70158151437481 0
    };
    \legend {positive, negative};
    \draw[dashed, very thick] (axis cs:0,2.4) -- (axis cs:6,2.4);
    \draw[dashed, very thick] (axis cs:0,3.1) -- (axis cs:6,3.1);
    \draw [draw=none, fill=boxcolor!20] (0.05,2.4) rectangle (6.15,3.1);
    \draw[dashed, very thick] (axis cs:3.4,0.0) -- (axis cs:3.4,2.1);
    \draw[dashed, very thick] (axis cs:3.4,2.1) -- (axis cs:6,2.1);
    \draw [draw=none, fill=boxcolor!20] (3.4,0.05) rectangle (6.5,2.1);
    \node[draw=none] at (5.5, 3.8) {\large $B_1$};
    \node[draw=none] at (2.8, 0.7) {\large $B_2$};
    \end{axis}
    \node[treenode] (dtree) at (6, 0.24\linewidth) {\yfeature $\le 3.1$}
    child { node[treenode] {\yfeature $\le 2.4$} edge from parent [treearrow]
      child { node[treenode] {\xfeature $\le 3.4$}
      child { node[nonrareclass] {negative} }
      child { node[treenode] {\yfeature $\le 2.1$}
          child { node[rareclass] {positive} }
          child { node[nonrareclass] {negative} } 
        }
      }
      child { node[rareclass] {positive} }
    }
    child { node[nonrareclass] {negative} edge from parent [treearrow] };
\end{tikzpicture}
    \caption{Two multidimensional boxes learned by a decision branches model from~\cite{lulf2023fast} for a 2D feature space. Each box is defined via orthogonal splits along the axes, each of which can be in turn translated into range queries by simply following the path from the root to the positive leaves.}
    \label{fig:range_query}
\end{figure}
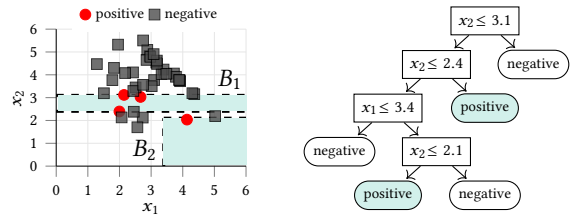

The query processing phase represents the actual procedure through which user queries are handled within our search engine. Given an incoming user query, defined by positive and negative instances, we gather the corresponding feature vectors for the data and train our decision branches model to separate these two classes from each other. The positive classes in the feature space are represented by multidimensional boxes (characterized by orthogonal boundaries along the axes) and can therefore be translated into range queries (in other words: SQL queries, in which predicates within the \texttt{WHERE} clause represent model decisions; see Figure~\ref{fig:range_query}). Our pre-built index structures enable efficient processing of these range queries, returning the target objects from the database during the inference phase. 
An essential characteristic of our decision branch classifier is its "index-awareness." This implies that the multidimensional boxes generated by the decision branches correspond to the same feature subsets covered by one of our pre-built index structures such that all queries can be answered by only using the index structures.

\section{Data set and Data Preprocessing}
\label{sec:dataset}
In our prototype, we showcase our search-by-classification approach on an aerial data set from Denmark of the year 2018 in a resolution of 12.5 cm per pixel. We split the data set into patches of the size 400x400 pixels, with a step size of 200 pixels. In total, this yielded $90,429,772$ patches that can be selected for various search tasks. In order to translate the image patches into meaningful feature vectors that can be used to train our decision branches, we used Vision Transformers (ViT)~\cite{vit} as feature extractor. We trained a ViT-T model using the self-supervised learning framework DINO~\cite{caron2021emerging} on a random subset of 400,000 patches of our aerial data set of Denmark for 100 epochs on one \emph{GPU RTX 3090}.\footnote{The feature quality and thereby also the quality of the query results can be even further improved by increasing the size of the training set as well as the model size.} 
By employing self-supervised training, the ViT model learns intrinsic data characteristics without being directed toward a specific classification task. This strategy ensures that the resulting features are versatile and applicable to an array of search tasks, rather than being confined to predefined classes.
After training the ViT, we extracted 384 features per patch from the final layer of the feature extraction model which in total consumed around 130 GB of storage for the entire data set.

\begin{figure*}
    \centering
    \input{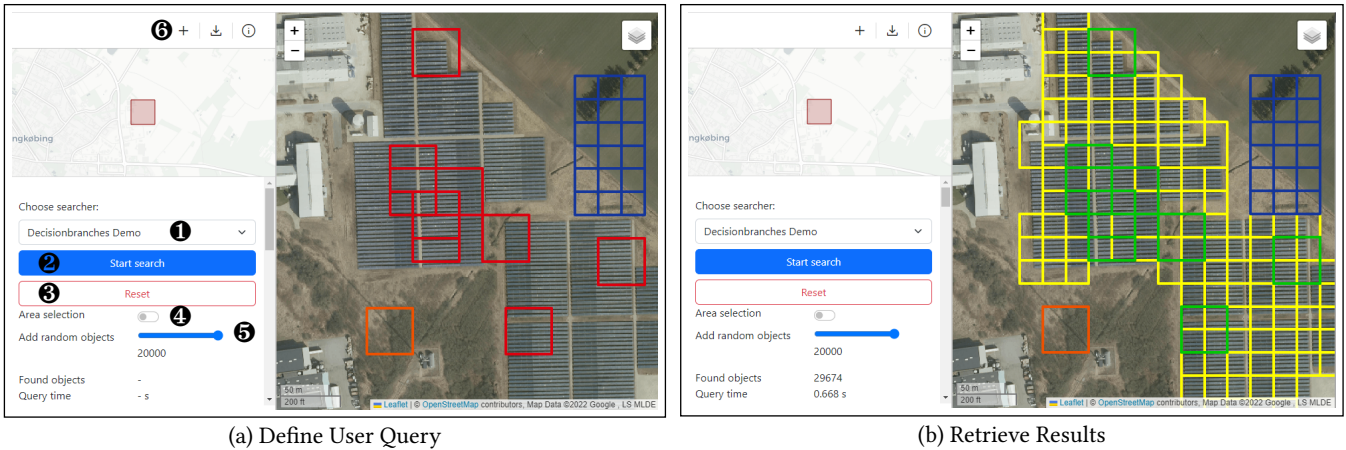}
    \vspace{-0.5cm}
    \caption{Visual search interface of \sysname. 
    The meanings of the buttons are described in Section \ref{sec:ui}.}
    \label{fig:screenshot}
\end{figure*}

\section{System Overview}
\label{sec:prototype}

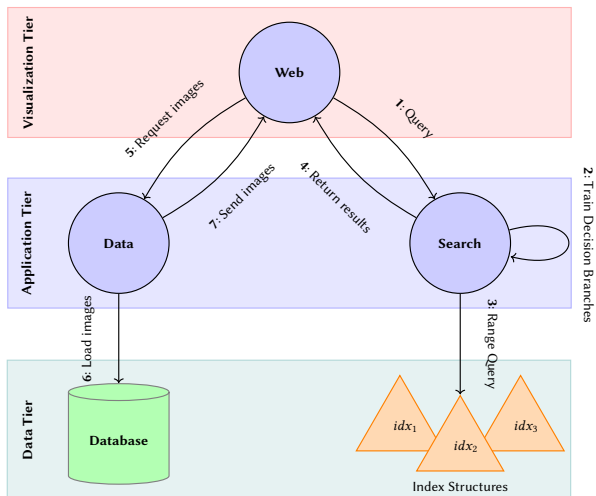
\begin{figure}[h]
    \centering    
    \scalebox{0.535}{%
        \hspace{0.6cm}
        \begin{tikzpicture}[
  node distance=6cm, 
  auto,
  minimum size=2.5cm,
  every node/.style={align=center, thick, font=\sffamily\large, fill=blue!20}, 
  database/.style={
    cylinder,
    fill = green!30,
    shape border rotate=90,
    aspect=0.25,
    draw=black!50,
    thick
  },
  triangle/.style={
    regular polygon,
    regular polygon sides=3,
    draw=orange,
    thick,
    fill=orange!30,
    text width=1em,
    inner sep=1mm
  },
]
  \node (web) [circle,draw] {\textbf{Web}};
  \node (data) [circle,draw, below left of=web] {\textbf{Data}};
  \node (search) [circle,draw, below right of=web] {\textbf{Search}};
  \node (db) [database, below of=data,node distance=4.9cm] {\textbf{Database}};  


\node (t1) [triangle, below=2cm of search, xshift=-1.5cm] {$idx_1$};  
\node (t3) [triangle, below=2cm of search, xshift=1.5cm] {$idx_3$};  
\node (t2) [triangle, below=2.5cm of search] {$idx_2$};  

  \node (index) [below of=t2, node distance=1cm, fill=none, draw=none] {Index Structures};

    \draw [->, thick, black] (web) to [bend left=15] node[midway, above, sloped, fill=none,yshift=-0.5cm] {\textbf{1}: Query} (search);
  \draw [<-, thick, black] (data) to [bend left=15] node[midway, above, sloped, fill=none,yshift=-0.5cm] {\textbf{5}: Request images} (web);
  \draw [<-, thick, black] (web) to [bend left=15] node[midway, below, sloped, fill=none,yshift=0.5cm] {\textbf{7}: Send images} (data);

  \draw [->, thick, black] (search) to [bend left=15] node[midway, below, sloped, fill=none,yshift=0.5cm] {\textbf{4}: Return results} (web);
      \draw [<-, thick, black] (db) to node[midway, above, sloped, fill=none,yshift=-0.5cm] {\textbf{6}: Load images} (data);

  \draw [->, thick, black] (search) to  node[midway, above, sloped, fill=none,rotate=180,yshift=-0.5cm] {\textbf{3}: Range Query} (t2);

\path [->, thick, black] (search) edge [loop right] node[midway, fill=none, sloped, xshift=-2cm, yshift=0.5cm] {\textbf{2}: Train Decision Branches} (search);

  \begin{pgfonlayer}{background}
    \node [draw=red!30, fit=(web) , inner sep=10pt,fill=red!10,
    minimum width=14cm, minimum height=3cm] (container1) {};
  \end{pgfonlayer}

  \begin{pgfonlayer}{background}
    \node [draw=blue!30, fit=(search) (data), inner sep=10pt,fill=blue!10, minimum width=14cm, minimum height=3cm] (container2) {};
  \end{pgfonlayer}
  
  \begin{pgfonlayer}{background}
    \node [draw=teal!30, fit=(db) (t1) (t2) (t3), inner sep=10pt,fill=teal!10,
    minimum width=14cm, minimum height=3cm,xshift=-0.5cm] (container3) {};
  \end{pgfonlayer}

   \node [align=left,xshift=-6.5cm,rotate=90,fill=none] at (container1.center) {\textbf{Visualization Tier}};
    \node [align=left,xshift=-6.5cm,rotate=90,fill=none] at (container2.center) {\textbf{Application Tier}};
 \node [align=left,xshift=-6.5cm,rotate=90,fill=none] at (container3.center) {\textbf{Data Tier}};


\end{tikzpicture}%
    }%
    \vspace{-0.3cm}
    \caption{Overview of \sysname's architecture.}
    \label{fig:systemarchitecture}
\end{figure}

The architecture of our search engine comprises three major components as shown in Figure \ref{fig:systemarchitecture}. These are:
\begin{enumerate}
    \item \textbf{Web application}: hosts the web frontend of the search engine where the user can browse the data and define the query intent. The frontend is implemented using JavaScript and the \emph{Leaflet}\footnote{\url{https://leafletjs.com/}} framework. 
    \item \textbf{Search application}: performs the actual search task by training a decision branch model on the query input and retrieves the corresponding database instances via range queries on the index structures. The search application is built with \emph{FastAPI}\footnote{\url{https://fastapi.tiangolo.com}}. For the decision branches as well as the index structures, we used the implementation available at \url{https://github.com/decisionbranches/decisionbranches}.
    \item \textbf{Data application}: stores the underlying image patches that are needed to show the results of the query (see Figure \ref{fig:demo_results}). The data application is also implemented as \emph{FastAPI} service.
\end{enumerate}

Before the search engine can be started, some preprocessing steps need to be taken: (a) The features for the image data need to be extracted (see Section \ref{sec:dataset}); (b) The index structures required for the range query processing must be built; (c) The lookup table mapping the image patches to their geolocations needs to be setup. Afterward, the search engine can be deployed and queries can be processed. In Figure \ref{fig:systemarchitecture}, the overall workflow of how a query is processed in our search engine is demonstrated. 
Once the user selects some positive and negative patches to create a user query and a decision branch model, the query information is sent to the search application~(Step 1). In the search application, the chosen model is trained with the labeled patch features, deriving range queries from the trained model~(Step 2).
These range queries are processed using the pre-built index structures~(Step 3). The resulting object IDs and their geolocations, along with certain query statistics such as the number of returned objects and query time, are then sent back to the web application~(Step 4).
Now, the web application starts to visualize the identified query rectangles on the web interface while concurrently requesting the corresponding image patches from the data application~(Step 5). These image patches for the discovered objects are fetched from the image database~(Step 6) and asynchronously loaded and displayed in the web interface sidebar~(Step 7).

\subsection{Search Models}
Our current prototype contains five different classification models for processing the user queries, namely \emph{Decision Branches}\footnote{Termed DBranch$_{[B]}$ in \cite{lulf2023fast}.}, \emph{Decision Branches Ensemble}\footnote{Termed DBEns$_{[B]}$ in \cite{lulf2023fast}.}  comprising 25 individual models, \emph{Decision Tree}, \emph{Random Forest} based on 25 decision trees and a \emph{Nearest Neighbor baseline}, where only the $1,000$ nearest neighbors are returned.
Note that only the decision branch models and the nearest neighbor baseline can leverage the pre-built index structures. The traditional scan-based approaches decision tree and random forest are included to highlight the difference in query response time of the utilized "index-aware" models in contrast to scan-based approaches. 
The nearest neighbor baseline is included for the sake of completeness. Although this model is based on a minor subset of the total 384 features, leading to a potential compromise on result accuracy, it does offer fast query response times. This efficiency is due to its utilization of the underlying k-d tree index structures.

\subsection{User Interface}
\label{sec:ui}
The web-based user interface of \sysname shown in Figure~\ref{fig:screenshot} and includes a configuration panel~(left) and a map panel~(right). These components work cohesively to ensure a seamless user experience.

\noindent\textbf{Configuration Panel.} \sysname includes a multitude of settings where users can: {\Large\ding{182}}~select the searcher, {\Large\ding{183}}~start the search for selected objects; {\Large\ding{184}}~reset the search to focus on a new class of objects; {\Large\ding{185}}~enable or disable the area selection feature, which enables users to select entire areas that to be labeled as negative (\eg, users could label entire forests to exclude them from the search); {\Large\ding{186}}~configure the number of negative samples that are added to the query; {\Large\ding{187}}~import and export user queries. Furthermore, the interface shows the images of the query results as well as the query statistics (query time and number of found objects). 

\noindent\textbf{Map Panel.} Transitioning to the right, we have an interactive web map that allows users to delve into geospatial and label data crucial for training the specified search models selected on the configuration panel. When zoomed to a specific level, users are able to select positive and negative instances that compose the training set---with a left mouse click marking positive instances (red rectangle) and a right click designating negative instances (blue rectangle). Note that the current position of the cursor is represented by an orange rectangle. Once a sufficient number of patches have been labeled, a simple click on the "Start Search" button~({\Large\ding{182}}) triggers the search. The search results are color coded to further assist the user, with green rectangles representing patches identified during the search but present in the training set, and yellow rectangles representing newly discovered ones.
\section{Demonstration}
\label{sec:demo}

\begin{figure}[t]
    \centering
    \input{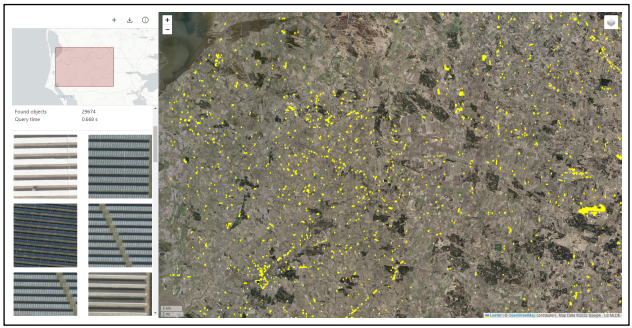}
    \caption{Visualization of search results on a map of Denmark. Yellow areas on the map correspond to found objects (\ie, solar panels). Individual search results are shown in the sidebar on the left. By clicking on a patch in the sidebar, the map panel automatically jumps to the corresponding patch’s location.}
    \label{fig:demo_results}
\end{figure}

In our demonstration, the audience can \begin{inparaenum}
    \item interact with \sysname to query objects over the country of Denmark, and
    \item refine posed queries based on the search results.
\end{inparaenum}
During the initial search, participants can query, \eg, solar panels. The search results are mapped for visual representation and individual patches are loaded onto the sidebar for closer examination (see Figure \ref{fig:demo_results}). These patches in the sidebar are arranged in descending order of the model's confidence. A higher position in the sidebar corresponds to more frequent appearances of the patches within the boxes of the decision branches.

One of the salient features of our search engine lies in its capacity for refinement. This empowers users to iteratively fine-tune their query based on the search outcomes. If the initial results aren't satisfactory, visitors will be invited to refine their queries by tweaking the positive and negative instances until the desired outcome is achieved. Unlike scan-based alternatives, which necessitate a full re-scan for each new query, our search engine deftly delivers new results within a span of seconds, optimizing user experience.\footnote{Underlying caching mechanisms of the OS can even further accelerate the adapted queries by caching already loaded parts of the index.}


\section{Summary}
\label{sec:summary}
We introduce \sysname as a tool to aid remote sensing and earth observation analysts in rapidly finding objects of interest in very large satellite imagery data sets. 
A short video showcasing demonstration scenarios is available at  \url{https://youtu.be/jwS96I1qhU8} and the source code for \sysname can be found at \url{https://github.com/decisionbranches/rapidearth}.
We encourage readers to adapt \sysname for their own data sets and use cases.

\begin{acks}
    Fabian Gieseke acknowledges support from the Independent Research Fund Denmark (grant number 9131-00110B) and from the German Federal Ministry of Education and Research (AI4Forest project; grant number 01IS23025).
\end{acks}


\bibliographystyle{ACM-Reference-Format}
\bibliography{literature}

\end{document}